\def\sorb{b}
\if s\sorb
  \documentstyle{article}
  \newlength{\absize}
  \renewcommand{\baselinestretch}{1.5}
  \renewcommand{\arraystretch}{0.5}
  
\begin{document}
  \date{}
  \pagestyle{empty}
  \thispagestyle{empty}
  \renewcommand{\thefootnote}{\fnsymbol{footnote}}
  \newcommand{\starttext}{\newpage\normalsize
    \pagestyle{plain}
    \setlength{\baselineskip}{4ex}\par
    \twocolumn\setcounter{footnote}{0}
    \renewcommand{\thefootnote}{\arabic{footnote}}}
\else
  \documentstyle[12pt,a4wide,epsf]{article}
  \newlength{\absize}
  \setlength{\absize}{\textwidth}
  \renewcommand{\baselinestretch}{2.0}
  \renewcommand{\arraystretch}{0.5}
  \begin{document}
  \thispagestyle{empty}
  \pagestyle{empty}
  \renewcommand{\thefootnote}{\fnsymbol{footnote}}
  \newcommand{\starttext}{\newpage\normalsize
    \pagestyle{plain}
    \setlength{\baselineskip}{4ex}\par
    \setcounter{footnote}{0}
    \renewcommand{\thefootnote}{\arabic{footnote}}}
\fi
\renewcommand{\theequation}{\thesection.\arabic{equation}}
\newcommand{\preprint}[1]{{\tt
  \begin{flushright}
    \setlength{\baselineskip}{3ex} #1
  \end{flushright}}}
\renewcommand{\title}[1]{%
  \begin{center}
    \LARGE #1
  \end{center}\par}
\renewcommand{\author}[1]{%
  \vspace{2ex}
  {\Large
   \begin{center}
     \setlength{\baselineskip}{3ex} #1 \par
   \end{center}}}
\renewcommand{\thanks}[1]{\footnote{#1}}
\renewcommand{\abstract}[1]{%
  \vspace{2ex}
  \normalsize
  \begin{center}
    \centerline{\bf Abstract}\par
    \vspace{2ex}
    \parbox{\absize}{#1\setlength{\baselineskip}{2.5ex}\par}
  \end{center}}

\setlength{\parindent}{3em}
\setlength{\footnotesep}{.6\baselineskip}
\newcommand{\myfoot}[1]{%
  \footnote{\setlength{\baselineskip}{.75\baselineskip}#1}}
\renewcommand{\thepage}{\arabic{page}}
\setcounter{bottomnumber}{2}
\setcounter{topnumber}{3}
\setcounter{totalnumber}{4}
\newcommand{\figsize}{}
\renewcommand{\bottomfraction}{1}
\renewcommand{\topfraction}{1}
\renewcommand{\textfraction}{0}
%%%%%%%%%%%%%%%%%%%%%%%%%%%%%%%%%%%%%%%%%%%%%%%%%%%%%%%%%%%%%%%%%%%%%%%%
% Arild's macros:
\newcommand{\beq}{\begin{equation}}
\newcommand{\eeq}{\end{equation}}
\newcommand{\beqa}{\begin{eqnarray}}
\newcommand{\eeqa}{\end{eqnarray}}
\newcommand{\nn}{\nonumber}

\newcommand{\dd}{\mbox{{\rm d}}}
\newcommand{\mH}{m_{\rm H}}
\newcommand{\dLips}{\mbox{{\rm dLips}}}

\def\Im{\mbox{\rm Im\ }}
\def\Re{\mbox{\rm Re\ }}
\def\fourth{\textstyle{1\over4}}
\def\gsim{\mathrel{\rlap{\raise 1.5pt \hbox{$>$}}\lower 3.5pt
\hbox{$\sim$}}}
\def\lsim{\mathrel{\rlap{\raise 1.5pt \hbox{$<$}}\lower 3.5pt
\hbox{$\sim$}}}
\def\GeV{{\rm GeV}}
%%%%%%%%%%%%%%%%%%%%%%%%%%%%%%%%%%%%%%%%%%%%%%%%%%%%%%%%%%%%%%%%%%%%%%%%
%
\def\slash#1{#1 \hskip -0.5em /}
%
%%%%%%%%%%%%%%%%%%%%%%%%%%%%%%%%%%%%%%%%%%%%%%%%%%%%%%%%%%%%%%%%%%%%%%%%%%
\def\Month{\ifcase\month\or
January\or February\or March\or April\or May\or June\or 
July\or August\or September\or October\or November\or December\fi}
\def\slash#1{#1 \hskip -0.5em /}
%%%%%%%%%%%%%%%%%%%%%%%%%%%%%%%%%%%%%%%%%%%%%%%%%%%%%%%%%%%%%%%%%%%%%%%%
%
% Here we go:
           
\preprint{University of Bergen, Department of Physics \\ 
Scientific/Technical Report No.\ 1995-02 \\ ISSN~0803-2696\\
hep-ph/9502283 \\
\Month, \the\year}

\vfill
\title{Testing $CP$ in the Bjorken process}

\vfill
\author{Arild Skjold \\ Per Osland \\\hfil\\
        Section for Theoretical Physics \\
        Department of Physics\thanks{Electronic mail addresses:
                {\tt \{skjold,osland\}@vsfys1.fi.uib.no}}\\
        University of Bergen \\ All\'egt.~55, N-5007 Bergen, Norway }
\date{}

%%%%%%%%%%%%%%%%%%%%%%%%%%%%%%%%%%%%%%%%%%%%%%%%%%%%%%%%%%%%%%%%%%%%%%%%
\vfill
\abstract{In a more general electroweak theory, there could be 
Higgs particles that are odd under $CP$, 
and also Higgs-like particles which are not eigenstates of $CP$. 
We discuss distributions which for the Bjorken process 
are sensitive to the $CP$ parity.
Correlations among  momenta of the initial electron and final-state 
fermions yield this kind of information.
We discuss also observables which may demonstrate presence of 
$CP$ violation and identify a phase shift $\delta$
which is a measure of the strength of $CP$ violation 
in the Higgs-vector-vector coupling,
and which can be measured directly in the decay distribution.
We present Monte Carlo data on the expected efficiency, and conclude that
it is relatively easy to determine whether the produced particle
is even or odd under $CP$. However, observation of any $CP$ violation 
would require a very large amount of data.}
  
\vfill

%%%%%%%%%%%%%%%%%%%%%%%%%%%%%%%%%%%%%%%%%%%%%%%%%%%%%%%%%%%%%%%%%%%%%%%%
\starttext

%%%%%%%%%%%%%%%%%%%%%%%%%%%%%%%%%%%%%%%%%%%%%%%%%%%%%%%%%%%%%%%%%%%%%%%%
\section{Introduction}
\label{sec:intro}
\setcounter{equation}{0}
%%%%%%%%%%%%%%%%%%%%%%%%%%%%%%%%%%%%%%%%%%%%%%%%%%%%%%%%%%%%%%%%%%%%%%%%
One of the main purposes of accelerators being planned and built
today, is to elucidate the mechanism of mass generation.
In the Standard Model mass is generated via an $SU(2)$ Higgs
doublet, associated with the existence of a Higgs particle,
whereas in more general models there are typically several
such Higgs fields, and also more physical particles.

Another fundamental issue is the origin of $CP$ violation.
While this question will be studied in considerable detail at
the SLAC B-Factory and at other dedicated $B$-physics
experiments, there is of course the possibility that $CP$ violation
may be related to the Higgs sector, as first suggested by
Weinberg \cite{Wein76}.
Therefore, when some Higgs candidate is discovered, it will be
important to determine it properties under $CP$.

In the context of Higgs production via the Bjorken 
mechanism~\cite{Bjorken}, 
we shall here consider how angular distributions may serve to disentangle 
a scalar Higgs candidate from a pseudoscalar one. In trying to probe the
uniqueness of the scalar character of the Higgs boson as provided by the
Standard Model, we have to confront its predictions with those
provided by possible extensions of the Standard Model.
Next, by allowing for $CP$ violation in the Higgs
sector, we briefly discuss some possible signals of such effects.
While the Standard Model induces $CP$ violation in the Higgs sector at 
the one-loop level provided 
the Yukawa couplings contain both scalar and pseudoscalar components
\cite{Wein}, we actually have in mind 
an extended model, such as e.g., the two-Higgs-doublet model
\cite{TDLee}.

Below we postulate an effective Lagrangian which contains $CP$ violation 
in the Higgs sector.
In cases considered in the literature, $CP$ violation usually 
appears as a one-loop effect. This is due to the fact that the $CP$-odd 
coupling introduced below
is a higher-dimensional operator and in renormalizable models these are
induced only at loop level. Consequently we expect the
effects to be small and the confirmation of presence of $CP$ violation 
to be equally difficult.
$CP$ non-conservation has manifested itself so far only in the neutral kaon
system. In the context of the Standard Model this $CP$ violation 
originates from the Yukawa sector via the CKM matrix \cite{CKM}. Although 
there may be several sources of $CP$ violation, including the mixing matrix, 
we will here consider a simple model where the $CP$ violation is restricted
to the Higgs sector and in particular to the coupling 
between some Higgs boson and the vector bosons. 
Specifically, by assuming that the coupling between the Higgs boson $H$ 
and the $Z$ has both scalar and pseudoscalar components, 
the most general coupling for the $HZZ$-vertex
relevant for the Bjorken process may be written as
\cite{Nel,Cha}
\beq
i\ 2^{5/4} \sqrt{G_{\rm F}}
\left[ m_Z^2 \ g^{\mu \nu}
+ \xi\left(k_1^2,k_2^2\right)
\ \left(k_{1} \cdot k_{2} \ g^{\mu \nu} - k_{1}^{\mu} k_{2}^{\nu} \right)
+ \eta\left(k_1^2,k_2^2\right)
\ \epsilon^{\,\mu \nu \rho \sigma} k_{1 \rho} k_{2 \sigma} \right],
\label{EQU:int1}
\eeq
with $k_{j}$ the vector boson momentum, $j=1,2$.
The first term is the familiar $CP$-even $Z^\mu Z_\mu \ H$ tree-level 
Standard Model coupling. 
The second term stems from the dimension-5 $CP$-even operator
$Z^{\mu\nu} Z_{\mu\nu} H$ with
$Z_{\mu \nu}=\partial_\mu Z_\nu -\partial_\nu Z_\mu$. The last term is $CP$ 
odd and originates from the dimension--5 operator 
$\epsilon^{\,\mu \nu \rho \sigma} Z_{\mu \nu} Z_{\rho \sigma} H$.
Simultaneous presence of $CP$-even and $CP$-odd terms leads to $CP$ violation,
whereas presence of only the last term describes a pseudoscalar 
coupling to the vector bosons.
The higher-dimensional operators are radiatively induced and we may therefore
safely neglect the contribution from the second term. This is due to the fact
that $CP$-violating effects always arise from interferences and since loops in
the Standard Model are already suppressed, we conclude that only 
new $CP$-violating effects that interfere with Standard Model tree amplitudes 
are potentially significant.
The strength parameter $\eta$ may in general be complex, with $\Im\eta$
describing the absorptive part of the amplitude arising from final-state 
interactions.

Related studies have been reported by \cite{Nel,Cha,Kniehl,Grzad} 
in the context of how to discriminate $CP$ eigenstates. However it should be
noted that our study takes advantage of the azimuthal angular distributions 
similar to the correlations between decay planes involving scalar and 
pseudoscalar Higgs bosons \cite{osskj}, including Monte Carlo data on the 
expected efficiency. In the context of $CP$ violation, related studies have
been reported by \cite{Cha}. 

\section{Distinguishing $CP$ eigenstates}
\label{sec:cpeig}
\setcounter{equation}{0}

We compare here the production of a Standard-Model Higgs ($h=H$) with
the production of a `pseudoscalar' Higgs particle ($h=A$) via the Bjorken 
mechanism,
\beq
e^{-}\left(p_{1}\right) e^{+}\left(p_{2}\right) 
\rightarrow f\left(q_{1}\right) {\bar f\left(q_{2}\right)} 
{h}\left(q_{3}\right).
\label{EQU:Bj1}
\eeq
The couplings of $H$ and $A$ to the vector bosons are given by retaining 
only the first and last term in (\ref{EQU:int1}), respectively.

Let the momenta of the two final-state fermions and the initial electron 
(in the overall {\it c.m.} frame) 
define two planes, 
and denote by $\phi$ the angle between those two planes
(see eq.~(\ref{EQU:Dj4}) below). 
Then we shall discuss the angular distribution of the cross section
$\sigma$,
\beq
\frac{1}{\sigma}\:
\frac{\dd\sigma}{\dd\phi}
\label{EQU:intro1}
\eeq
both in the case of $CP$-even and $CP$-odd Higgs bosons.

The fermion-vector couplings are given by $g_V$ and $g_A$.
As a para\-meterization of these, we define
the angles $\chi$ by
\beq
g_{V} \equiv g \cos \chi, \qquad
g_{A} \equiv g \sin \chi. 
\label{EQU:Dj9} 
\eeq
In the present work, the only reference to these angles 
is through $\sin2\chi$
(see table~1 of ref.~\cite{osskj}).
The differential cross section can then be written as
\beq
\dd^5\sigma_h
= \frac{G_{\rm F} N_1}{2\sqrt{2}\,s}\, 
 D(s,s_1)\, W_h \,
\dLips(s;q_1,q_2,q_3) , \qquad h = H, A,
\label{EQU:utgpkt} 
\eeq
with $\sqrt{s}$ the {\it c.m.} energy and 
$\dLips(s;q_1,q_2,q_3)$ denoting the Lorentz-invariant phase 
space. Furthermore, $N_1$ is a colour factor, which is three for 
quarks, and one for leptons. 
The momentum correlations are in the massless fermion approximation
given by
\beqa
W_H
& = & X_{+} - \sin 2 \chi \sin 2 \chi_{1} \ X_{-}, \nn \\
W_A
& = & \frac{|\eta(s,s_1)|^2}{m_Z^4}
\left[-2 X_{-}^{2}+ \frac{1}{4}\, s s_1 \left(Z_1
- \sin 2\chi \sin 2\chi_{1} Z_2 \right)\right],
\label{EQU:eqYA}
\eeqa
with $\sin 2 \chi$ and $\sin 2 \chi_{1}$ referring to the initial and final 
fermions, respectively, and where
\beqa
X_{\pm}
& = &
(p_1\cdot q_1)(p_2\cdot q_2) \pm(p_1\cdot q_2)(p_2\cdot q_1), \nn \\
Z_1
& = &
  [(p_1\cdot q_1)+(p_2\cdot q_2)]^2
+ [(p_1\cdot q_2)+(p_2\cdot q_1)]^2-\frac{1}{2}\, s s_1, \nn \\
Z_2
& = &
[(p_1+p_2)\cdot(q_1-q_2)][(p_1-p_2)\cdot(q_1+q_2)].
\eeqa
The normalization in eq.~(\ref{EQU:utgpkt}) involves the function 
\beq
D(s,s_1) = m_Z^4 \,
\frac{g_1^2}{(s_1-m_Z^2)^2+m_Z^2\Gamma_Z^2}\:
\frac{g_2^2}{(s-m_Z^2)^2+m_Z^2\Gamma_Z^2}\:, 
\label{EQU:defc} 
\eeq
with
$$s \equiv (p_1+p_2)^2,   \qquad
s_1 \equiv Q^2 \equiv (q_1+q_2)^2.  $$
Finally, $m_Z$ and $\Gamma_Z$ denote the mass and total width of
the $Z$ boson, respectively. 

We first consider angular correlations of two planes,
one spanned by the incident electron momentum (${\bf p_1}$) and
that of the final-state vector boson (${\bf Q}$), and the other
one spanned by the two final-state fermions (${\bf q_1}$ and ${\bf q_2}$).
Hence, we define the angle $\phi$ by
\beq
\cos \phi = \frac{\left({\bf {p}_{1}} \times {\bf {Q}}\right)
            \cdot \left({\bf {q}_{1}} \times {\bf {q}_{2}}\right)}
                      {|{\bf {p}_{1}} \times {\bf {Q}}| 
                       |{\bf {q}_{1}} \times {\bf {q}_{2}}|}.
\label{EQU:Dj4}
\eeq
Integrating the Higgs production cross section 
(\ref{EQU:utgpkt}) over the polar angle
of the vector boson (or Higgs) momentum, as well as over the
way the energy is shared between the two fermions, we find
\beqa
 \frac{\dd^2\sigma_h }{\dd\phi\: \dd s_1}
& = & \frac{N_1}{144 \sqrt{2} (4\pi)^4}\, \frac{G_{\rm F}}{s^2}
\sqrt{\lambda\left(s,s_1,m^2\right)}\, D(s,s_1) \, W_h^{\,\prime} ,
\qquad h = H, A, 
\label{EQU:no1} 
\eeqa
with azimuthal distributions given by the expressions
\beqa
W_H^{\,\prime}
& = &\lambda\left(s,s_{1},m^{2}\right)+12 s s_{1} 
+2s s_1 \cos 2 \phi \nn \\
& & +  \sin 2 \chi \sin 2 \chi_{1}  \left(\frac{3 \pi}{4}\right)^2 
\sqrt{s s_1}\, (s+s_1-m^2) \cos \phi,  \nn \\
W_A^{\,\prime}
& = &
\frac{|\eta(s,s_1)|^2}{m_Z^4} \lambda\left(s,s_{1},m^{2}\right)
2s s_1 \left(1-\frac{1}{4}\, \cos 2 \phi \right),
\label{EQU:no2} 
\eeqa
and where 
$\lambda\left(x,y,z\right)\equiv
x^{2}+y^{2}+z^{2}-2\left(x y + x z + y z\right)$
is the K\"allen function. 
The term $Z_2$ of eq.~(\ref{EQU:eqYA}) vanishes under the integration 
over the polar angle referred to above,       
and does not contribute in eq.~(\ref{EQU:no2}).
It would contribute to the forward-backward (with respect to 
the beam axis) asymmetry of the Higgs cross section.

\begin{figure}
\refstepcounter{figure}
\label{bjlepnlc}
\begin{center}
\mbox{\epsffile{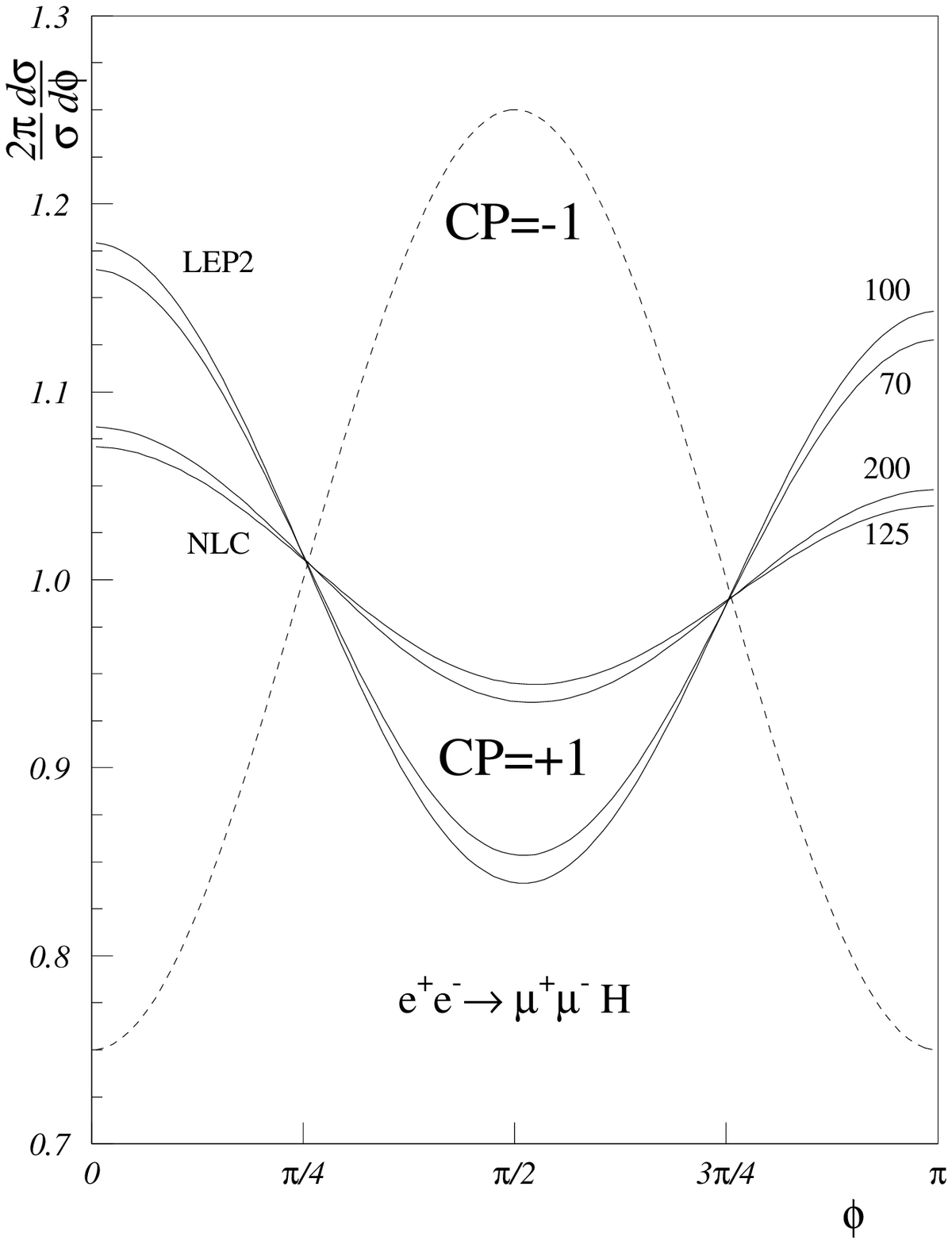}}
\vspace{20mm}
Figure~\thefigure
\end{center}
\end{figure}

A more inclusive distribution is obtained if we integrate
over the invariant mass of the final state fermion pair. 
Thus, let us consider
\beq
\frac{\dd{\sigma_h}}{\dd\phi}
= \int_{0}^{\left(\sqrt{s}-m\right)^2}\dd s_1 \:
\frac{\dd^2{\sigma_h}}{\dd\phi \: \dd s_1}. 
\label{EQU:Dk3}
\eeq
The distributions of eq.~(\ref{EQU:intro1}) take the form
\beqa
\frac{2 \pi}{\sigma_H}\:\frac{\dd\sigma_H}{\dd\phi} 
&=& 
1 + \alpha(s,m) \cos \phi 
  + \beta(s,m) \cos 2\phi, 
\label{EQU:Dl50} 
\\
\frac{2 \pi}{\sigma_A}\:\frac{\dd\sigma_A}{\dd\phi} 
&=& 
1 -\frac{1}{4} \cos 2\phi.
\label{EQU:Dl51} 
\eeqa
We shall consider the case when the energy is large enough to
allow both the Higgs and the $Z$ decaying to fermions to be on their mass
shells. We may then use the narrow-width approximation, effectively
setting $s_1=m_Z^2$, so that
\beqa
\alpha(s,m)
&=&
\sin2\chi\, \sin2\chi_1
\left(\frac{3\pi}{4}\right)^2\,
\frac{\sqrt{s}\, m_Z \left(s+m_Z^2-m^2\right)}
{\lambda\left(s,m_Z^2,m^2\right)+12 s\, m_Z^2}, \nn
\\
\beta(s,m)
&=&
\frac{2 s\, m_Z^{2}}{\lambda\left(s,m_Z^{2},m^{2}\right)
+12 s\, m_Z^{2}}.
\label{EQU:Dl53}  
\eeqa
At very high energies, $\lambda(s,m_Z^{2},m^{2})\sim s^2$,
and the coefficients $\alpha(s,m)$ and $\beta(s,m)$ will vanish
as $s^{-1/2}$ and $s^{-1}$, respectively.
Therefore, the Standard-Model distribution (\ref{EQU:Dl50}) will 
asymptotically become flat, whereas the
$CP$-odd distribution in eq.~(\ref{EQU:Dl51}) is independent 
of energy and Higgs mass.
A representative set of angular distributions is given 
in fig.~\ref{bjlepnlc} for the case 
$e^+e^- \rightarrow \mu^+\mu^-h$ for both LEP2 and higher energies,
and for different Higgs masses. 
(With $\phi$ being defined as the angle between
two oriented planes, it can take on values $0 \leq \phi \leq 2\pi$.)
Due to the $\sin2\chi$-facors in eq.~(\ref{EQU:Dl53}), 
$\alpha/\beta \simeq 0.1$ for the case of e.g. muons 
in the final state. This explains why the $\cos\phi$ contribution from
eq.~(\ref{EQU:Dl50}) is strongly suppressed in fig.~\ref{bjlepnlc}.
There is seen to be a clear difference between the
$CP$-even and the $CP$-odd cases. 

Experimentally, however, one faces the challenge of contrasting 
two angular distributions with a restricted number of events
and allowing also for background.
We shall here focus on the intermediate Higgs mass range; 
more specifically, we consider $m\lsim 140$~GeV where the Higgs 
decays dominantly to $b\bar{b}$. 
The main background will then stem from 
$e^+e^- \rightarrow ZZ$ and also 
$e^+e^- \rightarrow Z\gamma, \gamma \gamma$. 
The cleanest channel for isolating the Higgs signal from the background 
is provided by the $\mu^+\mu^-$ and
$e^+ e^-$ decay modes of the $Z$ boson. 

Let us next limit consideration to the energy range 
$\sqrt{s}=300-500$~GeV, as appropriate for a linear collider 
\cite{Wiik}, henceforth denoted NLC.
We impose the reasonable cuts and constraints described in 
\cite{Kniehl}; e.g. $|m_{\mu^+\mu^-}-m_Z| \leq 6$~GeV and 
$|\cos\theta_Z| \leq 0.6$,
where $m_{\mu^+\mu^-}$ denotes the invariant mass of the muon pair and
$\cos\theta_Z$
is the angle between ${\bf p_1}$ and ${\bf Q}$ given in eq.~(\ref{EQU:Dj4}).  
The signal for
$e^+e^- \rightarrow Z H \rightarrow \mu^+\mu^- b \bar{b}$ will then
be larger than the background
$e^+e^- \rightarrow Z Z \rightarrow \mu^+\mu^- b \bar{b}$
by an order of magnitude.
In the following we shall thus neglect the background
in the discussion of (\ref{EQU:Dl50}) versus (\ref{EQU:Dl51}).
With  $\sigma(e^+e^- \rightarrow Z H) \sim 200$~fb and an integrated 
luminosity of 20 ${\rm fb}^{-1}$ a year \cite{Kniehl}, 
about 4000 Higgs particles will be produced per year, in this
intermediate mass range. 
However, following~\cite{Kniehl} we have only $\sim 30$ signal events 
$e^+e^- \rightarrow Z H \rightarrow \mu^+\mu^- b \bar{b}$ left 
per year for e.g.\ a NLC operating at $\sqrt{s}=300$~GeV 
and a Higgs particle of mass $m=125$~GeV.
In the case $e^+e^- \rightarrow Z H \rightarrow e^+ e^- b \bar{b}$ 
we also have a t-channel background contribution 
from the $ZZ$ fusion process 
$e^+e^- \rightarrow e^+e^- (Z Z) \rightarrow e^+ e^- H$. 
This contribution may be neglected at LEP energies, 
but it is comparable to the s-channel contribution at higher
energies. However, this contribution can be suppressed
by imposing a cut on the invariant mass of the final-state electrons, 
e.g.\ $|m_{e^+e^-}-m_Z| \leq 6$~GeV.
Hence, we can effectively treat the electrons on the same footing as
the muons, thereby obtaining a doubling of the event rate.

Imposing the cut $|\cos\theta_Z| \leq b$, the predictions for 
the azimuthal correlations of eqs.~(\ref{EQU:Dl50})--(\ref{EQU:Dl51}) 
get modified. For the $CP$-even case we find
\beqa
\alpha^{b}(s,m)
&=          &
\sin2\chi\, \sin2\chi_1
\left(\frac{3\pi}{4}\right)^2\,
\frac{\sqrt{s}\, m_Z \left(s+m_Z^2-m^2\right)}
{\xi(b)\lambda\left(s,m_Z^2,m^2\right)+12 s\, m_Z^2}
\,\zeta(b),
\nn \\
\beta^{b}(s,m)
&=          &     
\frac{2\xi(b)s\, m_Z^{2}}
{\xi(b)\lambda\left(s,m_Z^{2},m^{2}\right)+12 s\, m_Z^{2}},
\label{EQU:El53}  
\eeqa
with
\beqa
\xi(b)
&=          &
\frac{1}{2}\left(3-b^2\right), \qquad \xi(1)=1, \nn \\
\zeta(b)
&=          &
\frac{2}{\pi}
\left(\frac{\frac{\pi}{2}-\arccos b}{b}+\sqrt{1-b^2}\right),
\qquad \zeta(1)=1, 
\eeqa
whereas for the $CP$-odd case
\beqa
-\frac{1}{4}
&\rightarrow&
-\frac{\xi(b)}{3+b^2}.
\label{EQU:El54}  
\eeqa

\begin{figure}
\refstepcounter{figure}
\label{bjmcnlc}
\begin{center}
\mbox{\epsfxsize=18cm\epsffile{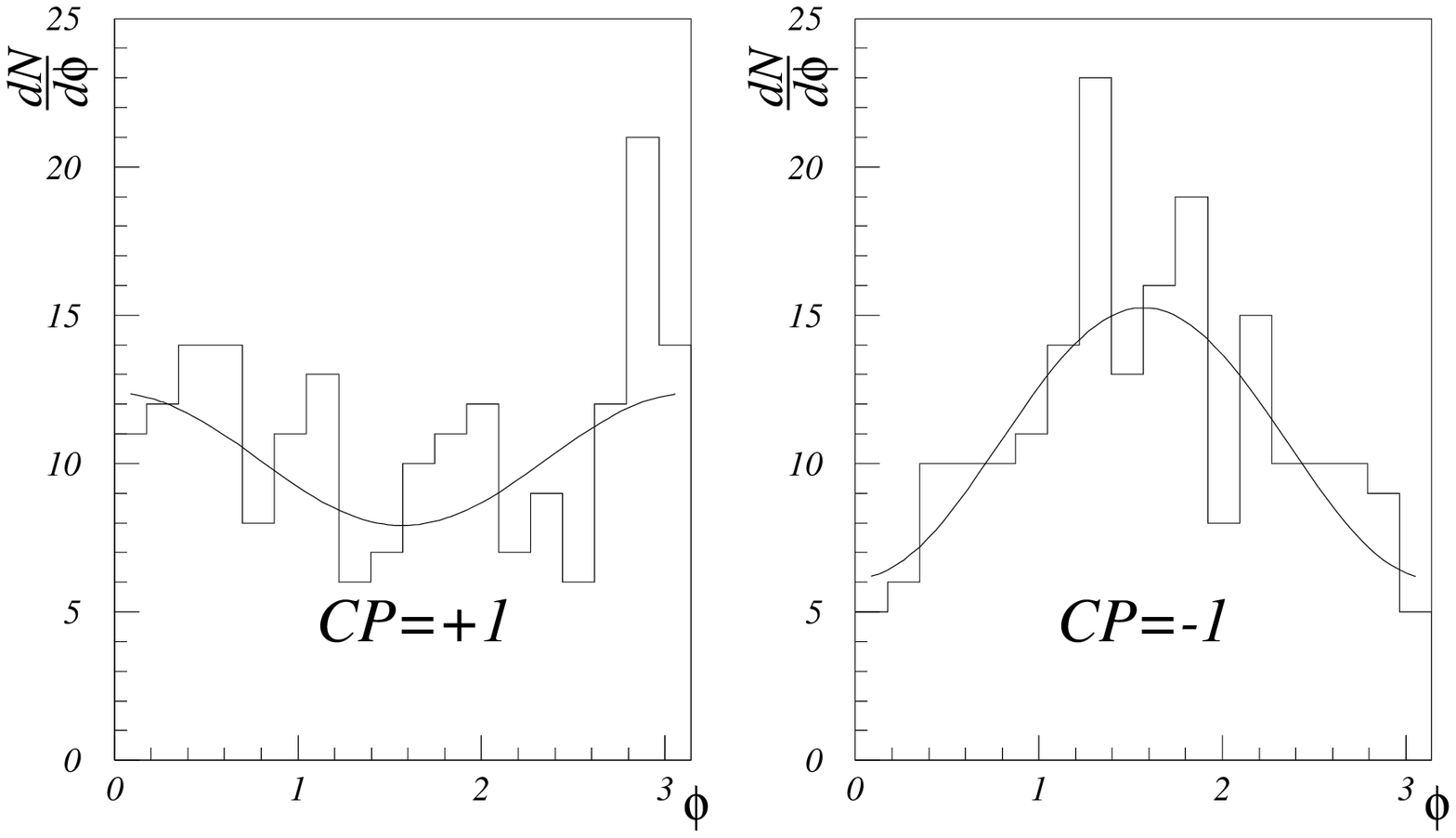}}

\vspace{20mm}

Figure~\thefigure
\end{center}
\end{figure}

In order to demonstrate the potential of the NLC for determining
the $CP$ of the Higgs particle,
we show in fig.~\ref{bjmcnlc} the result of a Monte Carlo simulation.
For this purpose we have used PYTHIA \cite{Sjostrand},
suitably modified to allow for the $CP$-odd case.
The statistics correspond to 3~years of running\footnote{The event
rate is based on the Standard Model, and could be different for
a non-standard Higgs sector.}
using both the $\mu^+\mu^-$ and $e^+ e^-$ decay modes of the $Z$ boson. 
This yields about 200 events in these channels.
As already stated, the $\alpha$ in (\ref{EQU:Dl53}) is small, 
and although the cut $b=0.6$ makes $\alpha$ increase as shown in 
(\ref{EQU:El53}), the $\cos\phi$ term is still too small to show up
in the Monte Carlo simulation.
For $\sqrt{s}=300~\GeV$ and $m_H=125~\GeV$, 
the `bare' prediction (\ref{EQU:Dl53}) for $\beta$ is 0.12, 
but the cut $b=0.6$ increases it slightly to 0.14. 
Similarly, the `-1/4' of (\ref{EQU:Dl51}) 
changes significantly to -0.39. Consequently, the cut makes it easier 
to discriminate between the $CP$-even distribution and the $CP$-odd one.
From fig.~\ref{bjmcnlc} we see that the individual angular 
Monte Carlo distributions are consistent with the predictions,
showing that a three-year data sample is large enough to
reproduce the azimuthal distributions.
In the Standard-Model case the fit gives $0.92\pm 0.07$ and 
$0.2\pm 0.1$ for the predictions 1.00 and 0.14, respectively, 
with $\chi^2=1.0$.
In the $CP$-odd case the fit gives $0.94\pm 0.07$ and 
$-0.4\pm 0.1$ for the predictions 1.00 and $-0.39$, respectively, 
with $\chi^2=0.7$.
More importantly, it is possible to verify the scalar nature of the
Standard-Model Higgs after about 3 years of running at the NLC since the
coefficient of the $\cos 2\phi$ term is more than 4 standard deviations away
from the corresponding coefficient for the $CP=-1$ case.
Using likelihood ratios, as described in \cite{Roe}, for choosing between the
two hypotheses of $CP$ even and $CP$ odd, we find that less than 3 years of
running suffices if we require 
a discrimination by four standard deviations.

\begin{figure}
\refstepcounter{figure}
\label{bjspec}
\begin{center}
\mbox{\epsfxsize=18cm\epsffile{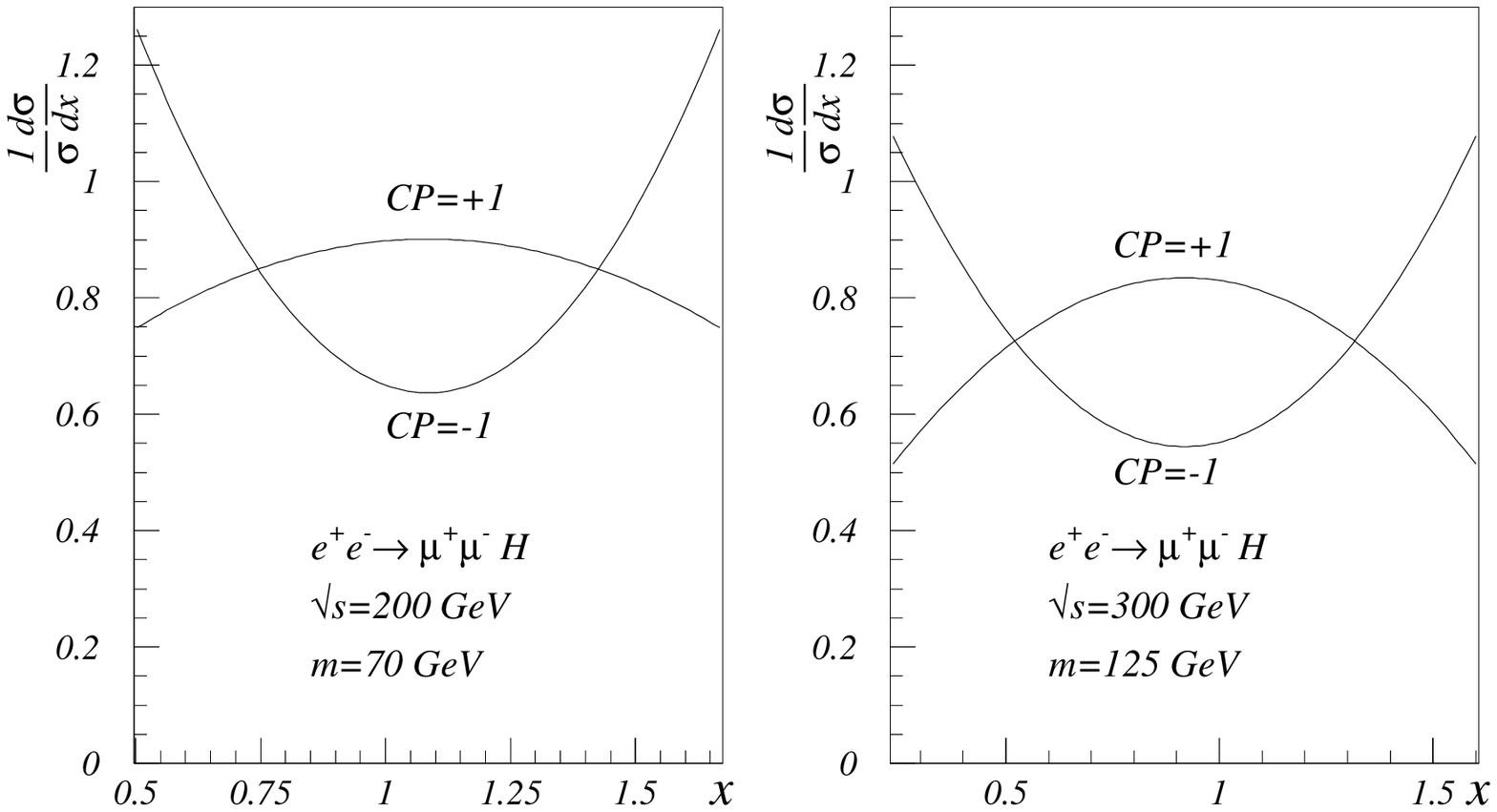}}

\vspace{20mm}

Figure~\thefigure
\end{center}
\end{figure}

An alternative test has recently been suggested 
by Arens et.~al.~\cite{Arens} in the context of Higgs 
decaying via vector bosons to four fermions, where one studies
the energy spectrum of one of the final-state fermions. 
Applying this idea to the Bjorken process one would
study the energy distribution of an outgoing fermion, e.g.\ $\mu^-$ or $e^-$. 
Introducing the scaled lepton energy, $x=4E_{l^-}/\sqrt{s}$, $l=\mu,e$,
we shall consider the energy distribution of the cross section
with respect to this final-fermion energy,
\beq
\frac{1}{\sigma}\:
\frac{\dd\sigma}{\dd x}
\label{EQU:intro2}
\eeq
both in the case of $CP$-even and $CP$-odd Higgs bosons. In the narrow-width
approximation we find 
\beqa
\frac{1}{\sigma_H}\:
\frac{\dd\sigma_H}{\dd x}
&=&
\frac{3 s^2}{4 \sqrt{\lambda}
\left(\lambda+12 s\, m_Z^2\right)}
\left[4 m_Z^2 +2\left(s+m_Z^2-m^2\right)x-sx^2\right], 
\label{EQU:Al52}
\\
\frac{1}{\sigma_A}\:
\frac{\dd\sigma_A}{\dd x}
&=&
\frac{3 s^2}{8 \lambda^{3/2}}
\left[\left(2\frac{\lambda}{s}+4 m_Z^2\right)-
2\left(s+m_Z^2-m^2\right)x+sx^2\right],
\label{EQU:Al53}
\eeqa
where $\lambda=\lambda\left(s,m_Z^2,m^2\right)$. 
The range in $x$ is given by $x_-\le x\le x_+$, with
\beq
sx_{\pm}=s+m_Z^2-m^2\pm\sqrt{\lambda}
\eeq
In this case there is a non-trivial dependence on the {\it c.m.} 
energy and the Higgs mass, also for the $CP$-odd case.
A representative set of energy distributions is given in 
fig.~\ref{bjspec} for the case $e^+e^- \rightarrow \mu^+\mu^- h$ 
for both LEP2 and NLC energies.
There is seen to be a clear difference between the $CP$-even 
and the $CP$-odd cases. 
Before we turn to the Monte-Carlo simulations, we shall impose the cut 
$|\cos\theta_Z| \leq b$, as in the case of angular correlations. 
This cut modifies eq.~(\ref{EQU:Al52}) so that
\beqa
\frac{1}{\sigma^b_H}\:
\frac{\dd\sigma^b_H}{\dd x}
&=&
\frac{3 s^2}{2 \sqrt{\lambda}
\left[\xi(b)\lambda+12 s\, m_Z^2\right]}
\Biggl\{2 m_Z^2\left(b^2-\frac{3 s\,m_Z^2}{\lambda}\left(1-b^2\right)\right)
\nn
\\
&+& 
\left[\left(s+m_Z^2-m^2\right)x-\frac{s}{2}x^2\right]\left(\xi(b)+
\frac{3 s\,m_Z^2}{\lambda}\left(1-b^2\right)\right)\Biggr\}, 
\label{EQU:Al54}
\eeqa
whereas the $CP$-odd distribution is independent of any cut in 
$\cos\theta_Z$. 
Of course the total cross section scales with $b$.

\begin{figure}
\refstepcounter{figure}
\label{bjmcx}
\begin{center}
\mbox{\epsfxsize=18cm\epsffile{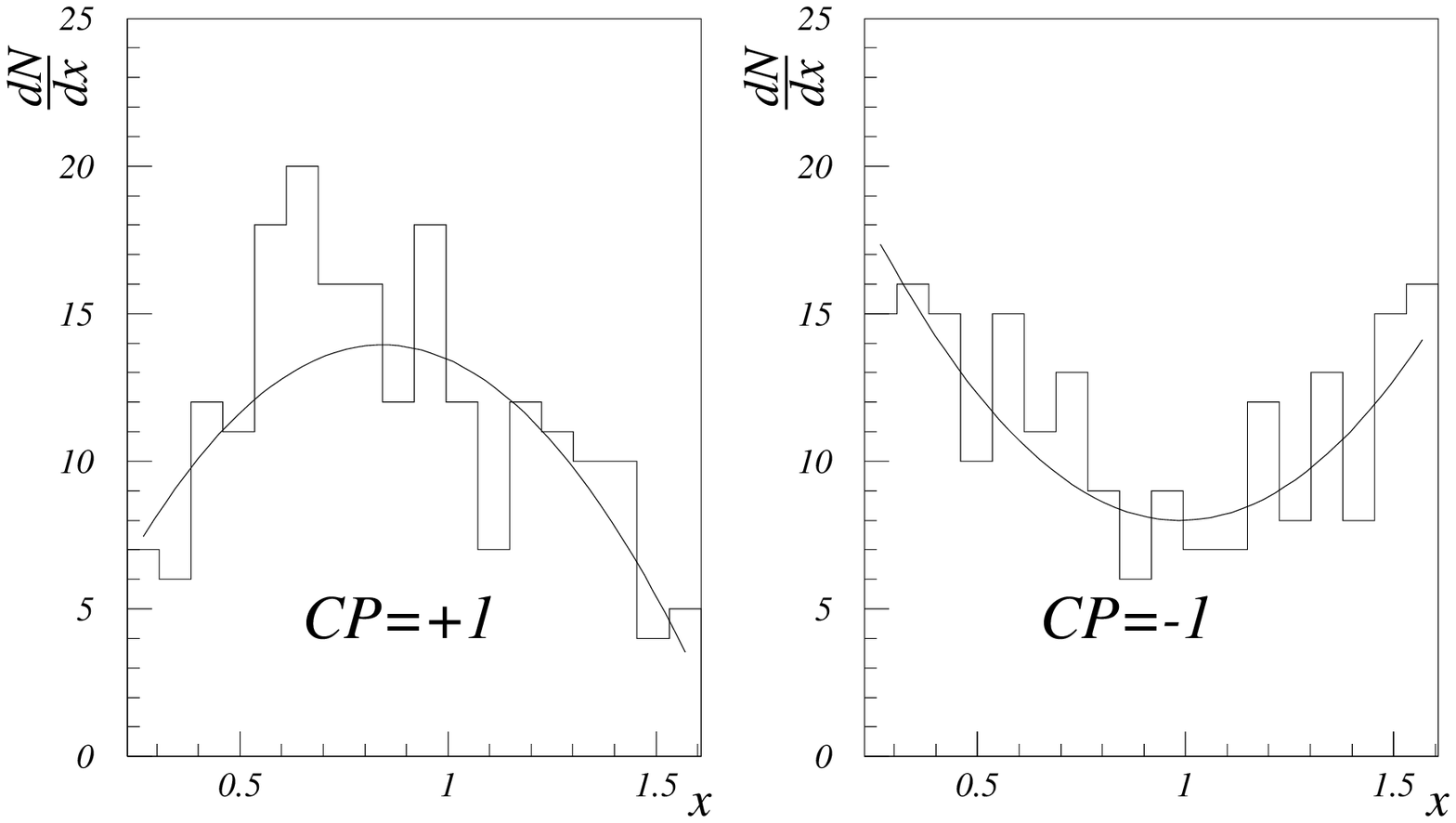}}

\vspace{20mm}

Figure~\thefigure
\end{center}
\end{figure}

In fig.~\ref{bjmcx} we show the result of a Monte-Carlo simulation 
for the energy distribution eq.~(\ref{EQU:intro2}) analogous 
to the one in fig.~\ref{bjmcnlc}. 
For $\sqrt{s}=300~\GeV$ and $m_H=125~\GeV$, 
the coefficients in (\ref{EQU:Al52}) and (\ref{EQU:Al53})
are $0.3, 1.3, -0.7$ and $1.5, -2.1, 1.1$, respectively, for increasing
powers of $x$. If we impose the cut $|\cos\theta_Z| \leq 0.6$, the
Standard-Model predictions are changed to $-0.003, 2.0, -1.1$. 
Hence, as in the case of angular distributions, 
the cut makes it easier to discriminate between the
$CP$-even distribution and the $CP$-odd one. 
In the Standard-Model case the fit gives $1.7\pm 0.2$ and 
$-0.9\pm 0.1$ for the predictions 2.0 and $-1.1$, respectively, 
with $\chi^2=1.0$. 
Naturally, the fit is not sensitive to the first coefficient.
In the $CP$-odd case the fit gives $1.6\pm 0.3$, $-2.2\pm 0.7$, and 
$1.1\pm 0.4$ for the predictions $1.5, -2.1$ and 1.1, respectively, 
with $\chi^2=0.6$. Also in this case a three-year data sample is enough to 
reproduce the predicted energy distributions. 
An analysis of the likelihood ratios demonstrates that less than 
3 years of running is sufficient if we require the correct answer 
with a discrimination by four standard deviations,
but more events seem to be required than in the case of angular
distributions. 

\section{$CP$ violation}
\label{sec:vio}
\setcounter{equation}{0}

As previously mentioned, if we allow for both the Standard-Model
and the $CP$-odd term in the Higgs-vector coupling (\ref{EQU:int1}),
then there will be $CP$ violation.
This situation will be discussed here.
It is similar to the case of Higgs decay discussed elsewhere
\cite{skjoldosland}.
We discard the higher-dimensional $CP$-even term for the reasons stated 
in the Introduction.

The differential cross section can then be written as
[cf.\ (\ref{EQU:utgpkt})--(\ref{EQU:defc})]
\beq
\dd^5\sigma
= \frac{G_{\rm F} N_1}{2\sqrt{2}\,s}\, 
 D(s,s_1)\left[ W_H+\frac{\Re \eta}{m_Z^2} W_1+\frac{\Im\eta}{m_Z^2} 
W_2+W_A \right]
\dLips(s;q_1,q_2,q_3). 
\label{EQU:utgpktcp} 
\eeq
The new momentum correlations are in the massless-fermion approximation
given by
\beqa
W_1
& = & -\epsilon_{\alpha\beta\gamma\delta}\,
p_{1}^{\alpha}p_{2}^{\beta}q_{1}^{\gamma}q_{2}^{\delta} \
\left[
Y_{-} - \sin(2\chi) \sin(2\chi_{1})Y_{+} \right], \nn \\
W_2
& = & \left(\sin 2\chi + \sin 2\chi_{1} \right) Y_1
-\left(\sin 2\chi - \sin 2\chi_{1} \right) Y_2,
\eeqa
where
\beqa
Y_{\mp}
& = &
(p_1\mp p_2)\cdot(q_1\mp q_2), \nn \\
Y_1
& = &
\left[\left(p_2 \cdot q_1 \right)-\left(p_1 \cdot q_2 \right)\right]
\left[\left(p_1 \cdot p_2 \right)\left(q_1 \cdot q_2 \right)
    +\left(p_1 \cdot q_2 \right)\left(p_2 \cdot q_1 \right)
    -\left(p_1 \cdot q_1 \right)\left(p_2 \cdot q_2 \right)\right], \nn \\
Y_2
& = &
\left[\left(p_1 \cdot q_1 \right)-\left(p_2 \cdot q_2 \right)\right]
\left[\left(p_1 \cdot p_2 \right)\left(q_1 \cdot q_2 \right)
    -\left(p_1 \cdot q_2 \right)\left(p_2 \cdot q_1 \right)
    +\left(p_1 \cdot q_1 \right)\left(p_2 \cdot q_2 \right)\right].
\eeqa
The term $W_2$ of eq.~(\ref{EQU:utgpktcp}), like $Z_2$ of
eq.~(\ref{EQU:eqYA}), vanishes under integration over the
polar angle. 

The distribution corresponding to (\ref{EQU:no1})--(\ref{EQU:no2}) 
can be written compactly as
\beqa
 \frac{\dd^2\sigma }{\dd\phi\: \dd s_1}
&  =   & \frac{N_1}{144 \sqrt{2} (4\pi)^4}\, \frac{G_{\rm F}}{s^2}
\sqrt{\lambda\left(s,s_1,m^2\right)}\, D(s,s_1) \nn \\
&\times&
\biggl[ \lambda\left(s,s_1,m^2\right)+4 s s_1 \left(1+2 \rho^{2} \right) 
+2s s_1\,\rho^2\,\cos 2(\phi + \delta) \nn \\
&  +   &   \sin 2 \chi \sin 2 \chi_{1} \left(\frac{3 \pi}{4}\right)^2
\sqrt{s s_1}\, (s+s_1-m^2)\,\rho\,\cos (\phi + \delta) \biggr] + {\cal
O}\left(\left(\Im\eta\right)^2\right), 
\label{EQU:no3}
\eeqa
with a modulation function
\beq
\rho=\sqrt{1+\left(\Re\eta\right)^2\lambda\left(s,s_1,m^2\right)/(4m_Z^4)}, 
\label{EQU:rho}
\eeq
and an angle 
\beq
\delta=
\arctan\frac{\Re\eta(s,s_1)\sqrt{\lambda(s,s_1,m^2)}}{2m_Z^2}, \qquad
-\pi/2 < \delta < \pi/2,
\label{EQU:delta}
\eeq
describing the relative shift in the angular distribution 
of the two planes, due to $CP$ violation. 
This rotation vanishes at the threshold for producing 
a real vector boson (where $\lambda=0$) and,
even for a fixed value of $\Re\eta$, grows with energy
(because of the $\sqrt{\lambda}$-factor). 
As discussed in the Introduction, the contribution from 
terms of order $(\Im\eta)^2$ may safely be neglected. 
However, the compact result (\ref{EQU:no3}) is valid for any $\Re\eta$. 
We will comment on how to probe $\Im\eta$ later.

This relation (\ref{EQU:delta}) can be inverted to give for the 
$CP$-odd term in the coupling:
\beq
\Re\eta=\frac{2m_Z^2}{\sqrt{\lambda(s,s_1,m^2)}}\, \tan\delta.
\eeq
This result (\ref{EQU:no3}) is completely analogous to the one
encountered for the decay of Higgs particles, eq.~(12) of
\cite{skjoldosland}, if we interchange $\phi$ and $\pi-\phi$.

Above threshold for producing a real vector meson accompanying the
Higgs particle, we may integrate over $s_1$ in the narrow-width
approximation.
Imposing the cut $|\cos\theta_Z| \leq b$, the distribution of 
eq.~(\ref{EQU:intro1}) takes the compact form
\beq
\frac{2 \pi}{\sigma^b}\:\frac{\dd\sigma^b}{\dd\phi} 
= 1 + \alpha^{b\, \prime}(s,m) \, \rho \, \cos (\phi +\delta)
+ \beta^{b\, \prime}(s,m) \, \rho^{2} \, \cos 2\left(\phi +\delta\right),
\label{EQU:Dl5} 
\eeq
with 
\beqa
\alpha^{b\, \prime}(s,m)
&=&
\sin2\chi\, \sin2\chi_1\left(\frac{3\pi}{4}\right)^2 
\frac{\sqrt{s}\, m_Z \left(s+m_Z^2-m^2\right)\zeta(b)}
{\lambda\left(s,m_Z^2,m^2\right)
[\xi(b)+ \left(3+b^2\right)s\,\left(\Re\eta\right)^2/2 m_Z^2] 
+ 12 s\,m_Z^2}, \nn
\\
\beta^{b\, \prime}(s,m)
&=&
\frac{2 s\ m_Z^{2}\xi(b)}{\lambda\left(s,m_Z^{2},m^{2}\right)
[\xi(b)+\left(3+b^2\right)s\,\left(\Re\eta\right)^2/2 m_Z^2]
+ 12 s\, m_Z^{2}}, 
\eeqa
and $\rho$ and $\delta$ given by eqs.~(\ref{EQU:rho})
and (\ref{EQU:delta}), substituting $s_1=m_Z^2$.

Any $CP$ violation would thus show up as a ``tilt'' in the
azimuthal distribution, by the amount $\delta$.
The amount could be extracted from a measurement of either
of the ``odd'' coefficients $A'$ or $B'$ in
\beq
\frac{2 \pi}{\sigma^b}\:\frac{\dd\sigma^b}{\dd\phi} =
1+A\left(s,m\right) \cos \phi + B\left(s,m\right) \cos 2\phi
 +A^{\prime}\left(s,m\right) \sin \phi
 + B^{\prime}\left(s,m\right) \sin 2\phi
\label{EQU:nr5}
\eeq
along the lines suggested in \cite{skjoldosland}.

\begin{figure}
\refstepcounter{figure}
\label{plcp}
\begin{center}
\mbox{\epsffile{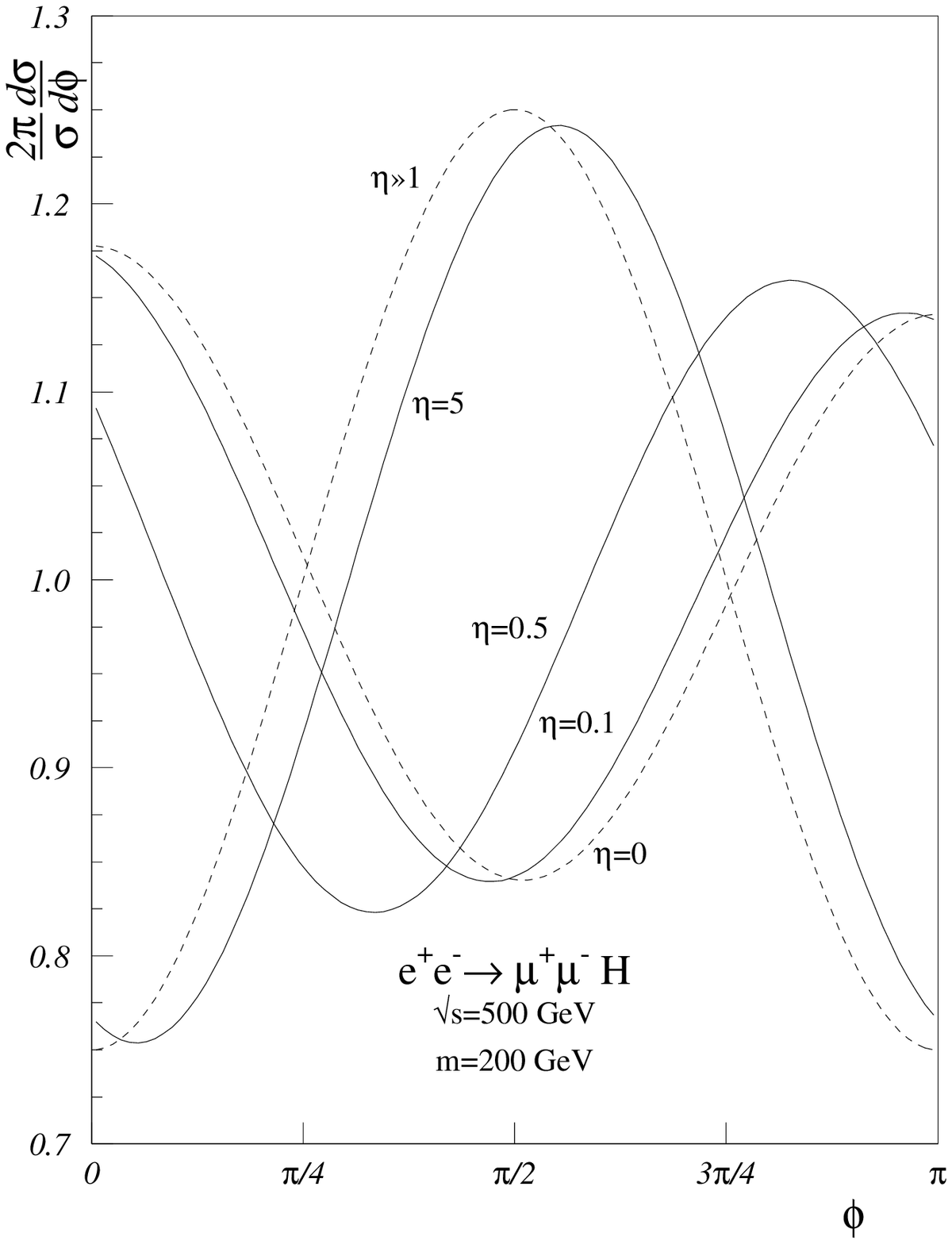}}

\vspace{20mm}

Figure~\thefigure
\end{center}
\end{figure}

A representative set of angular distributions is given in 
fig.~\ref{plcp} for a broad range of $\Re\eta$ values.  
We have considered a Higgs boson of $m=200$~GeV
accompanied by a $\mu^{+} \mu^{-}$-pair in the final state, 
produced at $\sqrt{s}=500$~GeV. 
We observe that for $\Re\eta \lsim 0.1$ and $\Re\eta \gsim 5$,
the deviations from the $CP$-even and $CP$-odd distributions,
respectively, are small. 
Experimentally it will be very difficult to disentangle two
distributions which differ by such a small phase shift. 
This should be compared with the situation 
in fig.~\ref{bjlepnlc} and fig.~\ref{bjmcnlc}.

We note that the special cases $\eta=0$ and $|\eta| \gg 1$ 
correspond to the $CP$ even and $CP$ odd eigenstates,
respectively. Hence, the distribution (\ref{EQU:Dl5})
should be interpreted as being intermediate between those for 
the two eigenstates; see fig.\ref{plcp}.                                    

In order to see how one can extract the dependence on the term 
proportional to $\Im\eta$, 
let us now turn to a discussion of energy asymmetries. 
We multiply the differential cross section eq.~(\ref{EQU:utgpktcp})
with the weighted energy difference 
$\left(\omega_1-\omega_2\right)/\left(\omega_1+\omega_2\right)$ for the two 
final fermions before integrating over energies. 
This energy-weighted differential cross section
corresponding to eq.~(\ref{EQU:no3}) takes the form
\beqa
 \frac{\dd^2\tilde{\sigma} }{\dd\phi\: \dd s_1}
& = & \frac{\Im\eta}{m_Z^2}
\frac{N_1 \sin 2 \chi_1}{36 \sqrt{2} (4\pi)^4}\, \frac{G_{\rm F}}{s}
\frac{\lambda^{3/2}\left(s,s_1,m^2\right)\, s_1 \, D(s,s_1)}
{s+s_1-m^2} 
\nn \\
&   & \times
\left[ 1+ 
\frac{\sin 2 \chi}{2 \sin 2 \chi_1}\left(\frac{3 \pi}{16}\right)^2
\frac{s+s_1-m^2}{\sqrt{s s_1}} \cos \phi\right], 
\label{EQU:no31}
\eeqa
since, in this case, only the $W_2$-term in (\ref{EQU:utgpktcp}) gives a 
non-vanishing contribution.
The energy-weighted differential cross section makes no reference neither 
to the $CP$-even nor to the $CP$-odd results,
but is proportional to $\Im\eta$ which describes
the absorptive part of the amplitude. 
A study of the above asymmetry thus allows us to probe for 
final state interactions and $CP$ violation in the Bjorken process.
 
%%%%%%%%%%%%%%%%%%%%%%%%%%%%%%%%%%%%%%%%%%%%%%%%%%%%%%%%%%%%%%%%%%%%%%%%
\section{Summary and concluding remarks}
\label{sec:conc}
\setcounter{equation}{0}

We have addressed the problem of estimating the amount of data needed
in order to distinguish a scalar Higgs from a pseudoscalar one 
at a future linear collider. 
We have argued that this is most likely not possible at LEP2.
However, we have demonstrated that one will be able to establish
the scalar nature of the Higgs boson at the Next Linear Collider 
from an analysis of angular and energy correlations. 
This study has been carried out for the case $\sqrt{s}=300$~GeV, 
$m=125$~GeV. Similar results are expected in other cases as long as 
the background is small.
In cases where the background can not be significantly suppressed a more
dedicated study would be required.

In order to establish or rule out specific models, one will also need
to compare different branching ratios, in particular to fermionic
final states.  The methods proposed above instead deal with quite
general properties of the models.

%%%%%%%%%%%%%%%%%%%%%%%%%%%%%%%%%%%%%%%%%%%%%%%%%%%%%%%%%%%%%%%%%%%%%%%%
%\section*{Acknowledgements}
It is a pleasure to thank Anne Grete Frodesen, Per Steinar Iversen, 
Conrad Newton and Torbj\"orn Sj\"ostrand for helpful discussions.
This research has been supported by the Research Council of Norway.
%%%%%%%%%%%%%%%%%%%%%%%%%%%%%%%%%%%%%%%%%%%%%%%%%%%%%%%%%%%%%%%%%%%%%%%%
                
%%%%%%%%%%%%%%%%%%%%%%%%%%%%%%%%%%%%%%%%%%%%%%%%%%%%%%%%%%%%%%%%%%%%%%%%
\renewcommand{\arraystretch}{1.5}
\clearpage
\centerline{\bf Figure captions}

\vskip 15pt
\def\fig#1#2{\hangindent=.65truein \noindent \hbox to .65truein{Fig.\ #1.
\hfil}#2\vskip 2pt}

\fig1{Angular distributions of the planes defined by 
incoming $e^-$ and final-state fermi\-ons for a $CP$-even Higgs particle 
(solid) compared with the corresponding distribution for a 
$CP$-odd one (dashed). 
Different energies and masses are considered in the $CP$-even case. 
We assume $\sqrt{s}=200$ and 500~GeV at LEP2 and NLC, respectively.
The considered values of the Higgs mass at the LEP2 are 70 and 100 GeV,
and at the NLC 125 and 200 GeV. In the $CP$-odd case there is no
dependence neither on energy nor on Higgs mass}

\fig2{Monte Carlo data displaying the angular distribution of events
$e^+e^- \rightarrow Z H \rightarrow l^+l^- b \bar{b}$, $l=\mu,e$ 
for a Standard-Model Higgs versus a CP-odd one. 
We have taken $\sqrt{s}=300$~GeV, $m=125$~GeV, and an angular cut
$|\cos\theta|\le b=0.6$.}

\fig3{Characteristic distributions for the scaled energy of the $l^-$,
$l=\mu,e$ in the Bjorken process $e^+e^- \rightarrow l^+l^- h$. 
Different energies and masses are considered.}

\fig4{Monte Carlo data displaying the lepton energy distribution for events
$e^+e^- \rightarrow Z H \rightarrow l^+l^- b \bar{b}$, $l=\mu,e$ 
for a Standard-Model Higgs versus a CP-odd one. 
We have taken $\sqrt{s}=300$~GeV and $m=125$~GeV. 
}

\fig5{Characteristic angular distributions for different amounts of $CP$
violation, including the $CP$-even ($\eta=0$) and $CP$-odd ($|\eta|\gg1$)
eigenstates. We have used
$\Re\eta=0.1, 0.5, 5$ for $\sqrt{s}=500$~GeV and $m=200$~GeV.}

%%%%%%%%%%%%%%%%%%%%%%%%%%%%%%%%%%%%%%%%%%%%%%%%%%%%%%%%%%%%%%%%%%%%%%%%

%\end{document}       % These two lines should be commented
%\endinput

%%%%%%%%%%%%%%%%%%%%%%%%%%%%%%%%%%%%%%%%%%%%%%%%%%%%%%%%%%%%%%%%%%%%%%%%
%\clearpage

%%%%%%%%%%%%%%%%%%%%%%%%%%%%%%%%%%%%%%%%%%%%%%%%%%%%%%%%%%%%%%%%%%%%%%%%
%\clearpage

%%%%%%%%%%%%%%%%%%%%%%%%%%%%%%%%%%%%%%%%%%%%%%%%%%%%%%%%%%%%%%%%%%%%%%%%
%\clearpage

%%%%%%%%%%%%%%%%%%%%%%%%%%%%%%%%%%%%%%%%%%%%%%%%%%%%%%%%%%%%%%%%%%%%%%%%
%\clearpage

%%%%%%%%%%%%%%%%%%%%%%%%%%%%%%%%%%%%%%%%%%%%%%%%%%%%%%%%%%%%%%%%%%%%%%%%
%\clearpage

%%%%%%%%%%%%%%%%%%%%%%%%%%%%%%%%%%%%%%%%%%%%%%%%%%%%%%%%%%%%%%%%%%%%%%%%

\begin{thebibliography}{99}

\bibitem{Wein76}
S. Weinberg, Phys.\ Rev.\ Lett.\ 37 (1976) 657.

\bibitem{Bjorken}
J.~D. Bjorken, in Proceedings of the 1976 SLAC Summer Institute 
on Particle Physics, ed.\ M. Zipf (SLAC Report No. 198, 1976) p.~22; \\
B.L.\ Ioffe and V.A.\ Khoze, Sov.~J.~Part.~Nucl.\ 9 (1978) 50; \\
D.R.T.\ Jones and S.T. Petcov, Phys.\ Lett.\ 84B (1979) 440; \\
J. Finjord, Physica Scripta 21 (1980) 143.

\bibitem{Wein}
S.\ Weinberg, Phys.\ Rev.\ Lett. 63 (1989) 2333; \\
Phys.\ Rev.\ D42 (1990) 860.

\bibitem{TDLee}
T.D. Lee, Phys.\ Rev.\ D8 (1973) 1226.

\bibitem{CKM}
N. Cabibbo, {\it Phys.\ Rev.\ Lett.}\ 10 (1963) 531;\\
M. Kobayashi and T. Maskawa, {\it Prog.\ Theor.\ Phys.}\ 49 (1973)
652.

\bibitem{Nel}
J.~R.\ Dell'Aquila and C.~A.\ Nelson, Phys.\ Rev.\ D33 (1986) 101. 

\bibitem{Cha}
D.\ Chang, W.~-Y.\ Keung and I.\ Phillips, Phys.\ Rev.\ D48 (1993) 3225.

\bibitem{Kniehl}
V. Barger, K. Cheung, A. Djouadi, B.A. Kniehl and P.M. Zerwas, \\
Phys.\ Rev.\ D49 (1994) 79.

\bibitem{Grzad}
B. Grz\c adkowski and J.F.\ Gunion, UCD preprint 95-5,
hep-ph/9501339.

\bibitem{osskj}
A. Skjold and P. Osland, Phys.\ Lett.\ B311 (1993) 261.

\bibitem{Wiik}
B.H.\ Wiik, in {\em HEP~93}, Proc.\ Int.\ Europhysics Conference
on High Energy Physics, Marseille, 22--28 July, 1993
eds.\ J. Carr and M. Perrottet
(Editions Frontieres, Gif-sur-Yvette, 1994) p. 739.
    
\bibitem{Sjostrand}
T. Sj\"ostrand, Computer Physics Commun.\ 82 (1994) 74.  \\
We use PYTHIA version 5.702 and JETSET version 7.404.

\bibitem{Roe}
B.P.\ Roe, {\it Probability and Statistics in Experimental Physics}, 
(Springer-Verlag Inc., New York, 1992).

\bibitem{Arens}
T.\ Arens, U.~D.~J.\ Gieseler and L.~M.\ Sehgal, 
Phys.\ Lett.\ B339 (1994) 127, \\
T.\ Arens and L.~M.\ Sehgal, PITHA 94/37, hep-ph/9409396, September 1994.
     
\bibitem{skjoldosland}
A. Skjold and P. Osland, Phys.\ Lett.\ B329 (1994) 305.

\end{thebibliography}
\end{document}